# An Assumption About The Free Energy Near The Critical Point

*Jianxiang Tian[#]*

*Department of Physics, Dalian University of Technology, Dalian116024, P.R.China*

*Department of Physics, Qufu Normal University, Qufu273165, P.R.China*

*Yuanxing Gui[*]*

*Department of Physics, Dalian University of Technology,Dalian116024, P.R.China*

*November 26[th], 2003*

**Abstract:** We divide the free energy near the critical point into two parts. One is the regular part, the other is the singular part. The singular part is assumed to be a concrete possible form. The singular part in this form is different from Widom scaling hypothesis and can give out the scaling laws of the critical exponents about the thermodynamic quantities. Furthermore, the critical exponents will be right worked out. We proof the singular part in this form can satisfy the qualities of the scaling function mentioned out by Widom. As an assumption, we expect the free energy in this form will be significative for the future research.

#: E-mail: lanmanhuayu@yahoo.com.cn.

*: E-mail: guiyx@dlut.edu.cn.

## 1. Introduction:

Experimental physicists discovered that the critical exponents about the thermodynamic quantities in various phase transitions satisfy the same scaling laws as:

$$\alpha + 2\beta + \gamma \approx 2 , \tag{1}$$

$$\gamma \approx \beta(\delta - 1). \tag{2}$$

When the ferromagnetic system is considered, $\alpha$ is the critical exponent of susceptibility, $\beta$ is the critical exponent of order parameter, $\gamma$ is the critical exponent of the specific heat, and $\delta$ is the critical exponent of the magnetic field coupled with order parameter. Their experimental values are [1]:

$$\alpha = 0.104 \pm 0.003 , \tag{3}$$

$$\beta = 0.325 , \tag{4}$$

$$\gamma = 1.23 , \tag{5}$$

$$\delta = 5.2 \pm 0.15 . \tag{6}$$

When Liquid-vapor phase transition is considered, $\alpha$ is the critical exponent of compressibility,

$\beta$ is the critical exponent of order parameter, $\gamma$ is the critical exponent of the specific heat, and $\delta$ is the critical exponent of the pressure field coupled with order parameter. Their experimental values are [2]:

$$\alpha = 0.110, \qquad (7)$$

$$\beta = 0.326, \qquad (8)$$

$$\gamma = 1.239, \qquad (9)$$

$$\delta = 4.80. \qquad (10)$$

L.Landau mentioned out the mean field theory in 1958 [3] and interpreted the phase transition phenomenon to some extent. In 1960 V.L.Ginzburg derived a very pleasing argument to show when the mean field theory would fail [4]. He argues that the mean field theory is correct when fluctuations in the order parameter are much smaller than its mean value. The result is near the critical point mean field theory is only reliable for dimensions greater than four [1].

In 1965, Widom supposed a singular part of the free energy of the system near the critical point behaves the form of scaling function and gave out the scaling laws above [5]. His work is normally called Widom Scaling Theory. Widom divided the free energy into two parts $F = F_r + F_s$. $F_r$ is the regular part. $F_s$ is the singular part. At the same time, $F_s$ is a scaling function $F_s(\lambda^p t, \lambda^q h) = \lambda^n F_s(t,h)$ with $h$ the proportional function of the field coupled with order parameter. In 1966, L.P.Kadanoff mentioned out scaling transformation [6] and gave out the physical interpretation of Widow Scaling Theory. In this paper, we assume a concrete possible form of the singular part of the free energy near the critical point. We proof this form leads to the scaling laws Eq. (1-2) and satisfies the qualities of scaling function. In the end we give out Fisher Scaling Law without the help of super-scale assumption.

## 2. Theory

### 2.1 Assumption

L.Landau suggested a most fruitful and general way of looking at mean field theory. When a ferromagnetic system in which the dimension of its order parameter is equal to one is considered, the free energy $F(T,M)$ near the critical point can be expanded in an exponent series by order parameter $M$ as:

$$F(T,M) = F(T,0) + \frac{1}{2}a(t)M^2 + \frac{1}{4}b(t)M^4 + \frac{1}{6}c(t)M^6 + \ldots, \qquad (11)$$

with

$$t = \frac{T - T_C}{T_C}. \qquad (12)$$

Order parameter $M$ is a tiny quality near the critical point. As we all know, the free energy in the form of Eq. (11) does not lead to right critical exponents.

As what Widom did, here we assume that the free energy near the critical point can be divided into two parts

$$F(M,t) = F_r(M,t) + F_s(M,t).$$  (13)

$F_s(M,t)$ behaves the form of

$$F_s(M,t) = D_1 t^y M + D_2 M^x,$$  (14)

with

$$D_1 = \begin{cases} D_{11} \neq 0, M > 0 \\ D_{12} \neq D_{11}, D_{12} \neq 0, M < 0 \end{cases},$$  (15)

$$D_2 = \begin{cases} D_{21} \neq 0, M > 0 \\ D_{22} \neq D_{21}, D_{22} \neq 0, M < 0 \end{cases},$$  (16)

$$x > 4,$$  (17)

$$y > 1.$$  (18)

$D_{11}$, $D_{12}$, $D_{21}$, $D_{22}$, $y$, and $x$ are constants to be assured. Eq. (14-15) is to satisfy the symmetry of the system for $M$ and $-M$ when $x$ and $y$ are assured.

*We do think that the singularity of about thermodynamic quantities near the critical point comes from $F_s(M,t)$. So it is acceptable to consider $F_r(M,t)$ as a constant $F_0$ when the singularity of these thermodynamic quantities near the critical point are considered.*

**2.2** $M > 0$

- **Scaling laws**

When $M > 0$, the free energy near the critical point is written as:

$$F = F_0 + \left( D_{11} t^y M + D_{21} M^x \right).$$  (19)

In the steady balanced state, the free energy has such qualities as

$$\frac{\partial F}{\partial M} = 0,$$  (20)

$$\frac{\partial^2 F}{\partial M^2} > 0.$$  (21)

Eq. (20) can be written as

$$D_{11} t^y + x D_{21} M^{x-1} = 0,$$  (22)

with the solutions

$$M = \left( -\frac{D_{11}}{x D_{21}} \right)^{\frac{1}{x-1}} t^{y/(x-1)}.$$  (23)

Thus,

$$M \propto t^{y/(x-1)} \ . \tag{24}$$

The magnetic field coupled with order parameter $M$ is

$$H = \frac{\partial F}{\partial M} = D_{11} t^y + x D_{21} M^{x-1} \ . \tag{25}$$

At the critical temperature $t = 0$,

$$H = x D_{21} M^{x-1} \ . \tag{26}$$

So,

$$H \propto M^{x-1} \ . \tag{27}$$

We calculate the susceptibility as follows

$$\begin{aligned}\chi &= \left(\frac{\partial M}{\partial H}\right)_T = \left(\frac{\partial H}{\partial M}\right)_t^{-1} \\ &= \left[x(x-1) D_{21} M^{x-2}\right]^{-1} \\ &= \frac{1}{x(x-1) D_{21}} M^{2-x}\end{aligned} \ . \tag{28}$$

When $H = 0$ and the steady balanced state is considered, the relation of $t$ and $M$ is assured by Eq. (23). Thus we have

$$\begin{aligned}\chi &= \frac{1}{x(x-1)D_{21}} \left[\left(-\frac{D_{11}}{xD_{21}}\right)^{\frac{1}{x-1}} t^{y/(x-1)}\right]^{2-x} \\ &= \frac{1}{x(x-1)D_{21}} \left(-\frac{D_{11}}{xD_{21}}\right)^{(2-x)/(x-1)} t^{[y/(x-1)](2-x)}\end{aligned} \ . \tag{29}$$

Thus we have

$$\chi \propto t^{[y/(x-1)](2-x)} \ . \tag{30}$$

The specific heat is defined as:

$$\begin{aligned}C_H &= T\left(\frac{\partial^2 F}{\partial T^2}\right)_H \\ &= T\left\{\frac{\partial}{\partial t}\left[\left(\frac{\partial F}{\partial t}\right)\left(\frac{\partial t}{\partial T}\right)\right]\frac{\partial t}{\partial T}\right\}_H \\ &= \frac{T}{T_c^2}\left(\frac{\partial^2 F}{\partial t^2}\right)_H \\ &= \frac{T}{T_c^2}\frac{\partial^2}{\partial t^2}\left(D_{11} t^y M + D_{21} M^x\right)\end{aligned} \ . \tag{31}$$

When $H = 0$ and the steady balanced state is considered, $C_H$ is solved to be:

$$C_H = \frac{T}{T_c^2} \frac{\partial^2}{\partial t^2} \left( D_{11} t^y M + D_{21} M^x \right)$$

$$= \frac{T}{T_c^2} \frac{\partial^2}{\partial t^2} \left\{ D_{11} t^y \left[ \left( -\frac{D_{11}}{xD_{21}} \right)^{\frac{1}{x-1}} t^{y/(x-1)} \right] + D_{21} \left[ \left( -\frac{D_{11}}{xD_{21}} \right)^{\frac{1}{x-1}} t^{y/(x-1)} \right]^x \right\} . \quad (32)$$

$$= \frac{T}{T_c^2} \left[ \left( -\frac{1}{x} \right)^{\frac{1}{x-1}} + \left( -\frac{1}{x} \right)^{\frac{x}{x-1}} \right] \left( \frac{yx}{x-1} \right) \left( \frac{yx-x+1}{x-1} \right) \left( D_{11}^{\frac{x}{x-1}} D_{21}^{\frac{1}{1-x}} \right) t^{\frac{yx}{x-1}-2}$$

$$\propto t^{\frac{yx}{x-1}-2}$$

The definitions of the critical exponents are [7]

$$\chi \propto |t|^{-\gamma}, \text{ when } H = 0, \quad (33)$$

$$C_H \propto |t|^{-\alpha}, \text{ when } H = 0, \quad (34)$$

$$M \propto |t|^{\beta}, \text{ when } H = 0, \quad (35)$$

$$H \propto M^{\delta}, \text{ when } t = 0. \quad (36)$$

Comparing Eq. (24, 27, 30, 32) with Eq. (33-36), we have such results as

$$\alpha = 2 - \frac{yx}{x-1}, \quad (37)$$

$$\beta = \frac{y}{x-1}, \quad (38)$$

$$\delta = x - 1, \quad (39)$$

$$\gamma = \frac{y}{x-1}(x-2). \quad (40)$$

Solving Eq. (37-40), we get Widom scaling law and Rushbrooke scaling law:

$$\alpha + 2\beta + \gamma = 2, \quad (41)$$

$$\gamma = \beta(\delta - 1). \quad (42)$$

- **Scaling transformations and scaling function**

We consider the transformations of the thermodynamic quantities as

$$t \to \lambda^p t, \quad (43)$$

$$H \to \lambda^q H, \quad (44)$$

$$M \to \lambda^m M, \quad (45)$$

$$F_s \to \lambda^s F_s. \tag{46}$$

Thus the transformation of Eq. (25) is

$$H = D_{11} t^y + x D_{21} M^{x-1}$$
$$\to$$
$$\lambda^q H = D_{11}(\lambda^p t)^y + x D_{21}(\lambda^m M)^{x-1}$$
$$\to$$
$$\lambda^q H = \lambda^{py} D_{11}(t)^y + x D_{21} \lambda^{m(x-1)}(M)^{x-1}. \tag{47}$$

It's easy to find that the dimensions of $\lambda^q H$, $\lambda^{py} D_{11}(t)^y$, and $x D_{21} \lambda^{m(x-1)}(M)^{x-1}$ are the same, and the dimensions of $H$, $D_{11}(t)^y$, and $x D_{21}(M)^{x-1}$ are the same. And it's also easy to find that

$$\lambda^q = \lambda^{py} = \lambda^{m(x-1)} \tag{48}$$

is a solution of Eq. (25, 47).
If Eq. (44) is right when arbitrary $\lambda$ is considered, we have

$$q = py = m(x-1). \tag{49}$$

Thus the transformation of Eq. (19) is

$$F_r = F - F_0 = \left(D_{11} t^y M + D_{21} M^x\right)$$
$$\to \lambda^s (F - F_0) = D_{11}(\lambda^p t)^y (\lambda^m M) + D_{21}(\lambda^m M)^x. \tag{50}$$
$$\to \lambda^s (F - F_0) = \lambda^{py+m} D_{11} t^y (M) + \lambda^{mx} D_{21} M^x$$

It is easy to find that

$$\lambda^s = \lambda^{py+m} = \lambda^{mx} \tag{51}$$

is a solution of Eq. (19, 50).
If Eq. (51) is right when arbitrary $\lambda$ is considered, we have
$$s = py + m = mx. \tag{52}$$

From Eq. (19, 50, 51), we know function $F_s = \left(D_{11} t^y M + D_{21} M^x\right)$ is a scaling function as:

$$\lambda^s F_s(t, M) = F_s(\lambda^p t, \lambda^m M). \tag{53}$$

- **Critical exponents**

Solving Eq. (49) and Eq. (52), we get

$$x = \frac{s}{s-q}, \tag{54}$$

$$y = \frac{q}{p}. \tag{55}$$

Inputting Eq. (54-55) to Eq. (37-40), we have

$$\alpha = \frac{2p - s}{q}, \tag{56}$$

$$\beta = \frac{s - q}{p}, \tag{57}$$

$$\delta = \frac{q}{s - q}, \tag{58}$$

$$\gamma = \frac{2q - s}{p}. \tag{59}$$

These results are just the ones of Widom's Scaling Theory. $s$ is normally interpreted to be the space dimension of the system.

### 2.3 $M < 0$

When $M < 0$ is considered, we replace $D_{11}$ and $D_{21}$ with $D_{12}$ and $D_{22}$ in the equations above where $D_{11}$ and $D_{21}$ appear. Eq. (24, 27, 30, 49, 37-42, 51- 59) will keep the same. We expect assumption Eq. (15-16) can bring the difference of the proportional coefficients in Eq. (23, 26, 29, 32) when $M < 0$ and $M > 0$ are considered separately in concrete physical models.

## 3. Discussion and conclusions

### 3.1 About Eq. (19)

From the analysis above, we have realized that the free energy in the form of Eq. (19) tends to good results including Widom's Scaling Laws and can obey Kadanoff's scaling transformation. The basis is that the transformations of the thermodynamic quantities obey the zoom-in ratio $\lambda^k$ with $k$ a constant and the assumption Eq. (14). The thing we should point out is that Eq. (14) is not a special case of Widom scaling hypothesis [8]: $F(T,H) = t^{1/j} \varphi(H/t^{l/j})$. Eq. (19) offers us information about the way to deal with the free energy from concrete physical model. We will discuss this problem in details in another paper, which is in process. When the liquid-vapor phase transition is considered, we choose $x = 5.80, y = 1.5648$, the critical exponents can be worked out by Eq. (37-40) to be

$$\alpha = 0.1092, \tag{60}$$

$$\beta = 0.326, \tag{61}$$

$$\delta = 4.80, \qquad (62)$$

$$\gamma = 1.2388. \qquad (63)$$

These data are very near to the experimental ones Eq. (7-10).

### 3.2 Correlation length $\xi$

We apply scaling assumption

$$F_s \propto \xi^{-d}. \qquad (64)$$

$d$ is the space dimension of the system. The steady balanced system is considered here. From Eq. (14, 32, 64), we can get

$$\xi \propto F_s^{-\frac{1}{d}} \propto \left( t^{\frac{yx}{x-1}} \right)^{-\frac{1}{d}} \propto t^{\frac{-yx}{x-1}\frac{1}{d}}. \qquad (65)$$

According to the definition of critical exponent $\upsilon$ [7]

$$\xi \propto t^{-\upsilon}, \text{ when } H = 0, \qquad (66)$$

we have

$$\upsilon = \frac{yx}{x-1}\frac{1}{d} = (2-\alpha)\frac{1}{d}. \qquad (67)$$

So,

$$\upsilon d = 2 - \alpha. \qquad (68)$$

This is Josephson Scaling Law.

### 3.3 Correlation function $G(r)$

Fisher modified Ornstein-Zernike Theory in 1964 and got the correlation function below [9]:

$$G(r)|_{T=T_C} \propto r^{-(d-2+\eta)}. \qquad (69)$$

Eq. (69) is the definition of critical exponent $\eta$ [7], too. $d$ is the space dimension. When a system in $d$ dimension space is considered, according to fluctuation theory, we have such result as [10, 11]

$$\chi \propto \int G(r)d\vec{r} \propto \int_{a_1}^{a_2} G(r)r^{d-1}dr. \qquad (70)$$

$a_1$ and $a_2$ are integral limits to be assured.

When the problem of phase transition is considered, there are three different important scales. They are crystal lattice constant $a$, correlation length $\xi(t)$, and space distance $r_e$. Here $a$ reflects material microscopic structure. $\xi(t)$ reflects the multi-body-action bound. $r_e$ reflects the effective bound of fluctuations in the order parameter in the theoretic description. When a macroscopic system is considered, these three scales satisfy such a relation as

$$a \leq r_e \leq \xi(t). \tag{71}$$

Thus we choose

$$a_1 = a, \tag{72}$$

$$a_2 = \xi(t). \tag{73}$$

Eq. (70) is assured to be

$$\chi \propto \int G(r)d\vec{r} \propto \int_a^{\xi(t)} G(r)r^{d-1}dr. \tag{74}$$

We operate Eq. (74) as follows:

$$\begin{aligned} \chi &\propto \int G(r)d\vec{r} \propto \int_a^{\xi(t)} G(r)r^{d-1}dr \\ &\rightarrow \frac{d}{dt}\chi \propto \frac{d}{dt}\int_a^{\xi(t)} G(r)r^{d-1}dr \end{aligned}. \tag{75}$$

Applying Eq. (29, 40, 66, 69), we get

$$[y/(x-1)](2-x)t^{[y/(x-1)](2-x)-1} \propto [G(\xi(t))](\xi(t))^{d-1}\frac{d}{dt}\xi(t) \propto t^{-\nu(2-\eta)-1}. \tag{76}$$

So,

$$t^{[y/(x-1)](2-x)-1} \propto t^{-\nu(2-\eta)-1}, \tag{77}$$

$$t^{-\gamma} \propto t^{-\nu(2-\eta)}. \tag{78}$$

Now we have

$$\gamma = \nu(2-\eta). \tag{79}$$

This is Fisher Scaling Law.

### 3.4 What we have done

In this paper we assume a concrete possible form of the singular part of the free energy near the critical point. The result is that we deduced scaling laws and the results of Kadanoff's scaling transformation successfully. In Part 3.3, Fisher Scaling Law Eq. (79) is given without the help of super-scale assumption firstly.